\documentclass[12pt,preprint]{aastex}
\received{2004 June 21}
\begin{document}
\def\etal{{\it et al.\/}}
\def\cf{{\it cf.\/}}
\def\ie{{\it i.e.\/}}
\def\eg{{\it e.g.\/}}
\def\n0{\dot{n}_\circ}
\def\d5{\dot{n}_{\circ,0.5}}


\title{Are all gamma--ray--bursts like GRB 980425, GRB 030329 and GRB 031203?}

\author{{\bf Dafne Guetta\altaffilmark{1}, 
Rosalba Perna\altaffilmark{2}, Luigi Stella\altaffilmark{3}
and Mario Vietri\altaffilmark{4}}} 

\altaffiltext{1}{Hebrew University of Jerusalem \&
Weizmann Institute, Israel.}
\altaffiltext{2}{Department of Astrophysical and Planetary Sciences,
University of Colorado at Boulder, Boulder, USA.}
\altaffiltext{3}{INAF-Osservatorio Astronomico di Roma, 
Monteporzio Catone (Roma), Italy.}
\altaffiltext{4}{Scuola Normale Superiore, Pisa, Italy}

\begin{abstract}

We study the probability that three GRBs (980425, 030329, 031203) are
found within $z=0.17$, given the luminosity functions consistent with
the $\log N-\log S$ relationship for classical cosmological bursts
(\ie, those observed by BATSE). We show that, in order for the
probability of these three low-$z$ events to be non-negligible (thus making
it more likely that they belong to the same class of the classical cosmological
bursts), the bursts' luminosity function must be a broken powerlaw.
By reasoning in analogy with beamed AGNs, we show that observations
are consistent with the expectations if GRB 980425 and GRB 031203 are
indeed normal bursts seen sideways. Within this model, no bright burst
within $z=0.17$ should be observed by a HETE--like instrument within
the next $\sim 20$ yr.

\end{abstract}


\section{Introduction}

After the discovery of the association of SN 2003dh
with  GRB 030329 (Stanek \etal, 2003; Hjorth \etal, 2003), 
and of SN 1998bw with
980425, it has been widely accepted that classical,
cosmological GRBs arise from the simultaneous collapse of a SN to
form a collapsar (Woosley 1994). Further evidence comes from the
third nearby burst, GRB 031203, which also has been seen to be
closely associated with a type Ic SN (Tagliaferri \etal, 2004;
Malesani \etal, 2004).
This model appears plausible for all long GRBs, even the distant ones,
in the light of the similarity between these two
objects and GRB 011121/SN2001ke (Bloom et al., 2002;
Garnavich et al. 2003), an object located at $z\simeq 0.36$. 
Still, the fact that only some bursts' afterglows display
the re--bumps now associated with the emergence of the optical
contribution due to the underlying SN beckons the question of
exactly which fraction of all GRBs (within the detectability
range of SN rebumps, say $z\lesssim 1$) are in fact associated with
simultaneous SNe. 
In order to tackle this problem, we study whether the detection
of GRB 980425, 030329
and 031203, the only ones unequivocally associated with a Type Ic
SN of extreme properties within a very small distance
(z=0.17) from us, is a statistical anomaly, for the luminosity functions
that are consistent with the GRB logN-logS distribution.

\section{Probabilities of detecting low-$z$ events}

If the three
low--redshift GRBs (980425, 030329 and 031203) are really typical
of the global GRB population, then their discovery 
within the current time and sky coverage, must be consistent with the
local GRB explosion rate as deduced from the very large GRB
samples made available by BATSE. A simple
computation of the expected number
of events within $z=0.17$ (the redshift of GRB 030329) 
shows that this is not the case. The volume out to $z=0.17$, in a
$\Omega_\Lambda =0.7, \Omega =1$ cosmology with $H_\circ =
71$ km s$^{-1}$ Mpc$^{-1}$, is $V = 1.4$ Gpc$^3$. Association of
a GRB with a SN requires an accurate position, and thus only
bursts revealed by the BeppoSAX WFC cameras, or HETE2 WXM telescopes,
can be used. Inclusion of INTEGRAL would only make matters slightly worse. 
These two X--ray telescopes monitor
$S_B= 0.123$ and $S_H =0.806$ steradians respectively (Band,
2002). The total effective operation times for BeppoSAX and HETE2
are generously estimated as $T_B = 4$ yr  and $T_H =2$ yr 
(L. Piro, private communication). If we call
$\n0$ the observed local rate of GRBs, we find that the total expected
number of GRBs located inside $z=0.17$ is
\begin{equation}
N =  \n0 V \frac{S_B T_B + S_H T_H }{4\pi} = 0.12 \frac{\n0}{0.5\;
 {\rm Gpc}^{-3} \;{\rm yr}^{-1}}
\end{equation}
The probability of observing three bursts, assuming
Poisson statistics, is then $P_{3} =
2.7\times 10^{-4} (\n0/(0.5\;
 {\rm Gpc}^{-3} \;{\rm yr}^{-1}))^3$. 

The major
uncertainty in the above equation concerns $\n0$, which takes on
different values depending upon exactly which data property is
fitted ($V/V_{max}$, or the $\log N-\log P$ relationship for BATSE
data). For the local observed rate,
Porciani and Madau (2001) consider three different
star formation rates, and correspondingly find three values for $\n0$, between
$0.11$ and $0.17$ bursts  Gpc$^{-3}$ yr$^{-1}$ (having converted
their results to H$_\circ$ = 71 km s$^{-1}$ Mpc$^{-1}$). Schmidt
(2001) considers the same star formation rates, but fits different
quantities, to obtain $\n0 = 0.48,0.51,0.72$ Gpc$^{-3}$ yr$^{-1}$.
Perna, Frail and Sari (2003) find $\n0 = 0.5$
Gpc$^{-3}$ yr$^{-1}$, while Guetta, Piran and Waxman (2003) find $\n0 =
0.44$ Gpc$^{-3}$  yr$^{-1}$. In the next section, 
we will repeat the analysis of Guetta \etal\/(2003) with a more updated
sample, to obtain $\n0 = 1.1$ Gpc$^{-3}$ yr$^{-1}$. 
With this value, the probability of a triple event would be 
$P_{3} = 2.9\times 10^{-3} (\n0/(1.1\; {\rm Gpc}^{-3}\; {\rm yr}^{-1}))^3$.
The hypothesis that the three nearby bursts,
980425, 030329 and 031203, belong to the same group as the
classical BATSE bursts, would thus be rejected with fairly high confidence.

However, this is not correct. The local rate discussed above has been
derived by all authors under the hypothesis that classical bursts
exceed by far the luminosity of GRB 980425 and they also exceed that
of GRB 031203 (i.e. they considered a minimum luminosity for the LF
$L_b\gtrsim 5\times 10^{49}$ erg/sec, which from now on we will assume
to define the minimum luminosity for the classical population).
Therefore, the derived probabilities are not self-consistent.  An
unified picture can only be obtained with a LF that includes all
luminosities down to that of GRB 980425, and at the same time yields
high probability of observing the three low-$z$ events.
This is what we will achieve in \S4, after setting up the formalism
and recomputing (for the more updated sample) the local rate for
the classical population in \S3. 

\section{Luminosity function for classical GRBs}

We consider all 2204  GRBs of the GUSBAD catalog which contains
all the long GRBs ($T_{90}> 2$sec) (Kouveliotou et al., 1993),
detected while the BATSE onboard trigger  (Paciesas et al. 1999)
was set for 5.5 $\sigma$ over background (resolution of 1024 ms)
in at least two detectors, in the energy range 50-300 keV.
We estimate $C_{\rm max}/C_{\rm min} $ for each
burst (where $C_{\rm max}$ is the count rate in the second
brightest illuminated detector and $C_{\rm min}$ is the minimum
detectable rate) and find $\langle V/V_{\rm max} \rangle=0.335
\pm 0.007$.

The luminosity function (LF) of the ``classical'' population
of long-duration GRBs may be represented as a power law with lower
and  upper limits, $L_{\rm b}$ and $L_{2}$, respectively. The
local LF of GRB peak luminosities $L$, defined as
the co-moving space density of GRBs in the interval $\log L$ to
$\log L + d\log L$ is:
\begin{equation}
\label{Lfun} \Phi_o(L)=c_o(L/L_{b})^{-\alpha} \;, \,\,\,\;\;  L_{\rm b}
< L < L_2
\end{equation}
where $c_o$ is a normalization constant. We stress that this
LF is the ``isotropic-equivalent" LF, it does not include 
the effects of beaming.

We assume that GRBs trace the star-formation history,
and adopt the recent SFR derived by Rowan-Robinson (1999; RR-SFR);
this can be fitted with the expression
$ R_{GRB}(z) = \n0 \, {\rm max}(0.75,0.75\,z)$

The modeling procedure involves the derivation of the peak flux
P(L,z) of a GRB of peak luminosity L observed at redshift z:
\begin{equation}
\label{peak} P(L,z)=\frac{L}{4\pi D_L^2(z)}
\frac{C(E_1(1+z),E_2(1+z))}{C(E_1,E_2)}\;,
\end{equation}
where $ D_L(z)$ is the bolometric luminosity distance and
$C(E_1,E_2)$ is the spectral energy distribution integrated
between $E_1 = 50$ keV and $E_2=300$ keV. Schmidt (2001) finds
that the median value of the spectral photon index in the
50-300keV band for long bursts GRBs is -1.6. We use this value for our
analysis to include the k-correction.

Objects with luminosity $L$ observed by BATSE with a flux limit
$P_{\rm lim}$ are detectable to a maximum redshift $z_{\rm
max}(L,P_{\rm lim})$  that can be derived from Eq. (\ref{peak}). 
We consider an average limiting flux $P_{\rm lim}=0.25$ ph cm$^{-2}$s$^{-1}$
taken from the BATSE catalog.

The number of bursts with a peak flux $>P$ is given by:
\begin{eqnarray}
 \nonumber N(>P)=\int\Phi_o(L)d\log L
\\  \int_0^{z_{max}(L,P)} \frac{R_{GRB}(z)}{1+z}
\frac{dV(z)}{dz}dz 
\end{eqnarray}
where the factor  $(1+z)^{-1}$ accounts for  cosmological time
dilation and $dV(z)/dz$ is the comoving volume element.

To be consistent with previous calculations,
we take the low luminosity cut-off $L_b\sim 5\times 10^{49}$ erg s$^{-1}$, 
while the high luminosity, $L_2\sim 5\times 10^{52}$  erg s$^{-1}$,
is taken on the order of the maximum luminosity detected until the time of writing 
(Bloom et al. 2003). The slope $\alpha$ 
of the LF is constrained by fitting the model with the
observed peak flux distribution with a non-linear
Levemberg-Marquart minimum $\chi^2$ method. To avoid problems with 
error correlation in a cumulative distribution like $N(>P)$ propagate in an
unknown way, we use the differential distributions $n(P)\equiv
dN/dP$ for this analysis.
We  find that the best fit  
is given for $\alpha\sim 0.72$; the p-value of the fit is 
$\sim 0.70$ (see Fig.1 upper panel). 
Both $\alpha$ and the normalization value
are somewhat insensitive to the choice of $L_b$ below a value
$\sim 10^{50}$ erg s$^{-1}$. This is  
mainly because GRBs with very low luminosity
appear above the sensitivity limit of $\sim 0.25$ photons cm$^{-2}$
s$^{-1}$ in a very small volume around the observer.

To obtain the observed local rate of GRBs per unit volume, $\n0$, we need
to  estimate the effective full-sky coverage of our GRB sample.
The GUSBAD catalog represents 3.185 years of BATSE full sky
coverage implying a rate of 692 GRB per year. Using our LF
 we find $\n0\sim 1.1 \,$Gpc$^{-3}$yr$^{-1}$.
\footnote[1]{Note that for a different SFR like for example the SF2-SFR of 
Porciani and Madau (2001) we still find a good fit and a local 
rate smaller by a factor $\sim 2$. We find a similar result also for the LF 
studied in the next session.}

\section{Luminosity function for bursts seen from any direction and revised probabilities}

We are thus left with the question of how to include GRB 980425
and GRB 031203 within our analysis. It is quite possible that these bursts
belong to a different, local population of bursts, as occasionally
claimed by some authors for GRB 980425 (Bloom et al. 1998). Still, there are at least two 
arguments suggesting that 980425 is just a normal burst seen sideways
(Nakamura 1998; Eichler \& Levinson 1999; Woosley, 
Eastman \& Schmidt 1999). One is the exceptional similarity of the 
two underlying SNe
in 980425 and 030329, within a class (that of extreme Type Ic SNe) known
instead for its lack of common patterns. The other is  that
GRB 021211 (della Valle \etal, 2003), and  all bursts with the
so--called optical rebumps are likely associated also to SN.

For this reason, we consider a LF model which includes GRBs seen sideways. 
We follow the analogy with beamed AGNs, 
according to which (Urry and Shafer 1984) the total
LF is simply a broken power--law, agreeing, at the bright end,
with the LF of the objects seen face--on. We thus take:
\begin{equation}
\label{broken}
\Phi_o(L)=c_o \left\{ \begin{array}{ll}(L/L_{\rm b})^{-\gamma},\;\;\; &
L_1 < L < L_{\rm b} \\ (L/L_{\rm b})^{-\alpha},\;\;\; & L_{\rm b} < L <
L_2\;.
\end{array}\right.
\end{equation}{}
We impose here $L_1 = L_{980425}$, $L_2 = 5\times 10^{52}$ erg s$^{-1}$ and
$L_{\rm b}$ 
to be the border value between GRBs seen sideways
(weak bursts) and the ones seen face--on (classical bursts). We choose
somewhat arbitrarily  $L_b\sim 5\times 10^{49}$ erg s$^{-1}$, and 
fit the same data as in the previous section
with $\alpha$ and $\gamma$ as free parameters.
Note that it has been proposed that X-ray flashes (XRFs) are simply GRBs 
viewed off-axis. If this interpretation is correct, then
XRFs could be included in our analysis in 
the lower part of the LF broken power law (recalling that XRFs are
generally underluminous compared to GRBs). 
The best--fit ($\alpha=0.7,\,
\gamma=0.1$ and p-value $\sim 0.76$) is shown in Fig. 1 (lower panel).
\footnote[2]{Note that the curve is almost identical to the one corresponding to the best fit LF 
obtained for the classical bursts shown in the upper panel. This is because
to add a low luminosity ($L<5\times 10^{49}$ erg/sec) tail to the LF
does not affect the logN-logS distribution (low luminosity
GRBs appear above the BATSE sensitivity limit in a reduced volume as explained in section 4).}
With this LF, the total local rate of events, down to
the lowest luminosity (that of 980425) is now much higher, $\n0 = 10$
Gpc$^{-3}$yr$^{-1}$,  
while the rate for $L>L_b$ remains unaffected. Using Eq(1), we 
find that the expected number of events within $z=0.17$, and of luminosity
even as low as GRB 980425, is $N_l=2.4$, which is clearly highly compatible
with the current observation of 3 events.

A strong radio emission is expected
from bursts seen sideways  at $\sim $ 1 yr delay 
(Waxman 2004ab, Livio \& Waxman 2000, see also Waxman 2004a,b 
and reference within) as the jet decelerates to 
sub-relativistic speed and its emission approaches isotropy.
However no late-time rebrightening was detected for either GRB 980425
or GRB 031203 (Sodeberg, Frail \& Wieringa 2004;
Sodeberg et al. 2003, 2004).
These results suggest that neither GRB 980425 nor GRB 031203 were off-axis
events, and instead were intrinsically sub-energetic GRBs.
Therefore we also consider the possibility that both GRB 980425 and GRB 031203 
might have intrinsically low luminosity and all GRBs are seen
on axis. 
The LF can still
have an {\em intrinsic} break for some (unknown) reason, and
therefore it can still be represented by eq.(\ref{broken}) 
(as in Schmidt 2001).
In this case we can leave $L_b$ as a free parameter in our fit 
and look for the best fit parameters $\alpha, \gamma, L_b$.
The results ($\alpha=0.95,\, \gamma=0.4,\, L_b=10^{51}$ erg s$^{-1}$
and p-value $\sim 0.64$)
are shown in Fig. 1 lower panel. With this LF the 
total local rate of events is  $\n0 = 20$ Gpc$^{-3}$  yr$^{-1}$.
However, if all GRBs are seen on axis,  then there is no 
natural explanation
for the break in the LF. Indeed a single power law LF 
with  $L_1 = L_{980425}$, $L_2 = 5\times 10^{52}$ erg s$^{-1}$
could be rapresentative of these bursts.
However in this case we get a very high local rate, $\n0\sim 100$ Gpc$^{-3}$ yr$^{-1}$,
which would predict about 24 events within the $z=0.17$ volume (again using Eq.1). The probability
of having only 3 occurrences is $\sim 10^{-7}$, which makes this scenario very unlikely.
Therefore, we conclude that a broken powerlaw  for the luminosity function is highly preferred to a single 
powerlaw, if the three low-$z$ events do belong to the same class as the classical ones.
Also note that the rate for the single powerlaw would yield about 400 GEM (galactic events
per million years), implying an implausibly large number of GRB
remnants (Loeb \& Perna 1998; Efremov et al. 1998) at any given time.

Our  model also allows us to compute the
probability of finding one object (GRB 980425) as close as $z=0.008$,
which would be of course prohibitively small if we were to apply
to this burst the statistics for classical 
GRBs discussed in \S3. For our best-fit powerlaw model,
we find that the probablity of bursts as luminous as
GRB 980425 within the volume accessible to 
BeppoSAX and HETEII, is $\approx 10^{-3}$, small but
still not completely ruled out.
For the detection threshold of Swift (Gehrels et al. 2004), 
this rate would be 0.05, while
that of all bursts with $L<L_b$ is $\sim 2$ for a 2-yr observation period.

\clearpage
\begin{figure}[t]
{\par\centering \resizebox*{0.65\columnwidth}{!}{\includegraphics
{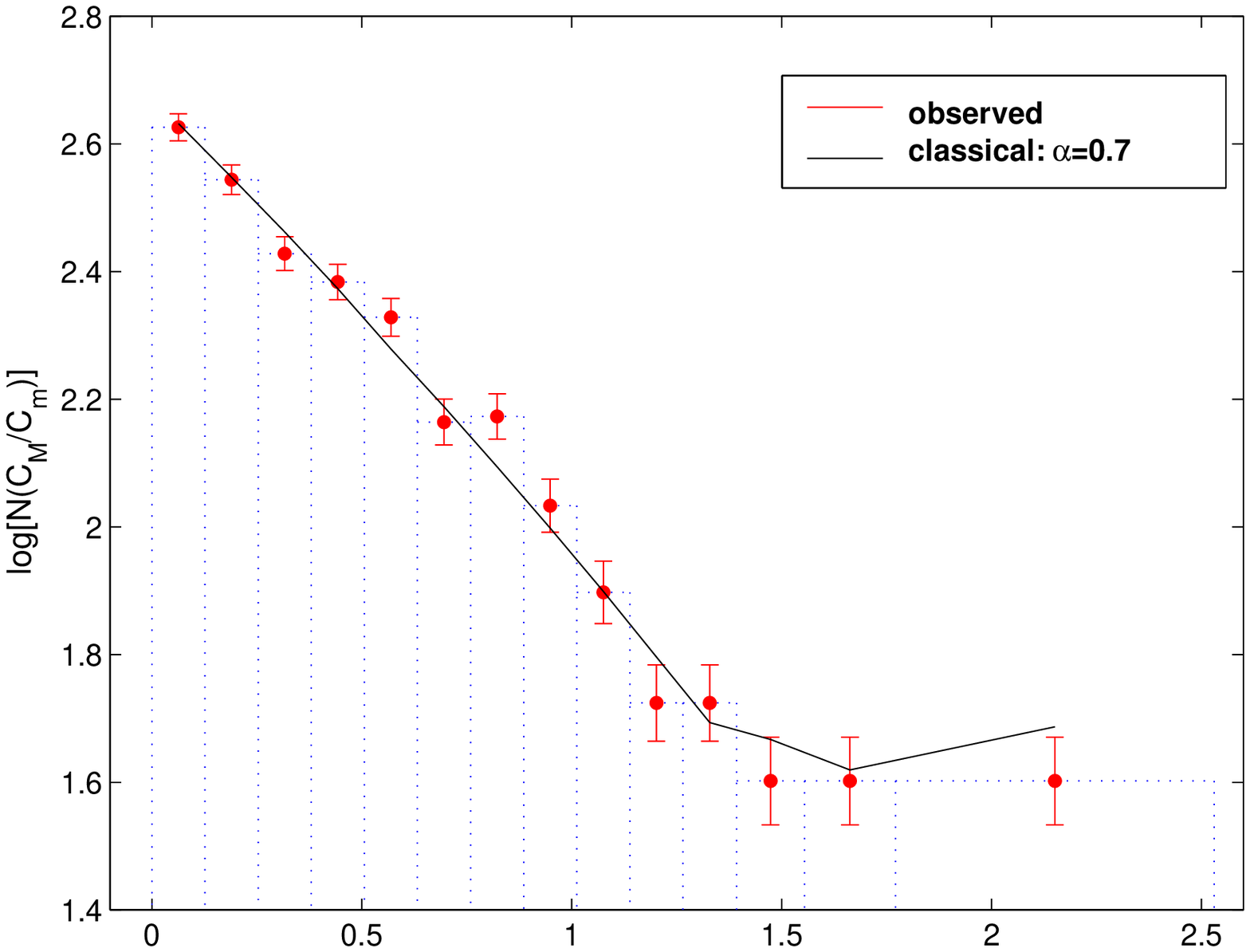}} \par}
{\par\centering \resizebox*{0.65\columnwidth}{!}{\includegraphics
{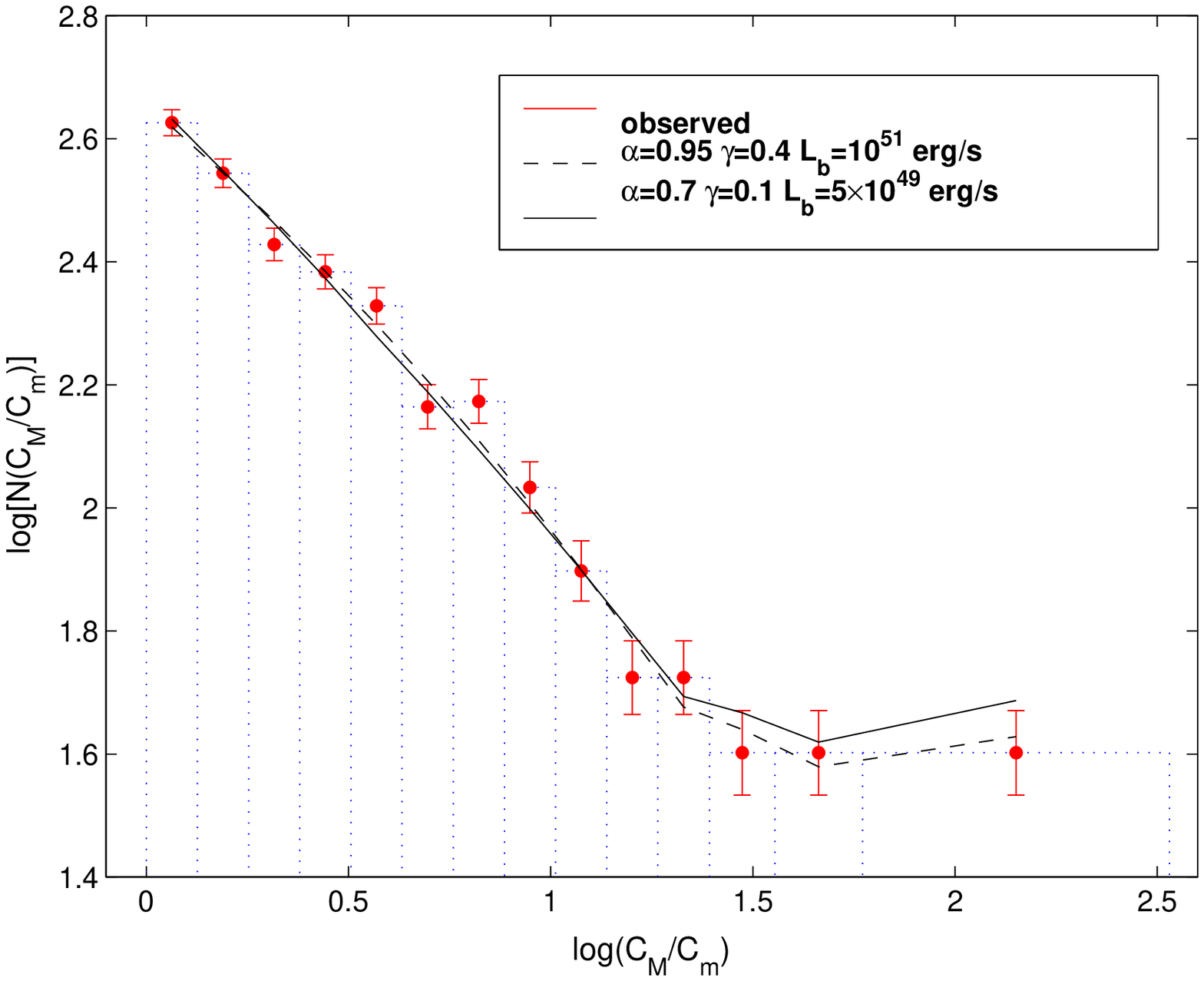}} \par}
 \caption{\label{fig2}
Upper panel: The best-fit 
differential distribution of the classical GRBs vs. the  observed
one taken from the GUSBAD catalog.
Lower panel: The best-fit
differential distribution of the GRBs in the case in which $L_b$
marks the border between off-axis and on axis-GRBs (solid line)
and in the case in which we leave $L_b$ as a free parameter  (dashed line)
vs. the observed one taken from the GUSBAD catalog. 
For both panels: The log$_{10}(C_M/C_m)$ is divided into 14 bins, the first 11 of equal
size and the remaining 3 with varying sizes chosen such that the number of bursts per bin
in the observed sample is at least $N_{\rm min}=40$ (in order to have reasonable statistics,
so that the Poisson error will not be too large), and N is the observed ({\it circles}) or 
theoretical ({\it lines}; only the values at the center of each bin count) number
of bursts in each bin. The edges of the bins are also plotted.}
\end{figure}
\clearpage

\section{Conclusions}

We have shown in this {\it Letter} that the presence of three
nearby GRBs does not pose a problem for the current view
(according to which both distant and nearby bursts belong to the
same class), if we modify the LF
to include an extension to luminosities as low as that of GRB 980425.
We did this by means of a {\it minimum impact} extension, \ie, by
assuming, in complete analogy with beamed AGNs, that the LF of
the bursts is a broken power--law, with the bright--end distribution
equal to what we derived when we neglected the presence of faint,
sideways bursts. As a by--product, we showed that this yields a
non--negligible probability for the detection of GRB 980425--like
event, which is also a new result.

How could this picture change? Some authors have insisted that it
is not possible to throw GRB 980425 in the same cage as classical
GRBs, because of its unique properties in the radio band. 
Within our model, the local rate of
nearby bright bursts ($L\ga 5\times 10^{49}$ erg s$^{-1}$) 
observable by HETEII within $z=0.17$ is $\n0
V S_H/4\pi = 0.057$ yr$^{-1}$. This implies that we ought to
observe the next such event $\sim 20 $ yr from now. Should we see one
significantly earlier than that then the argument
suggesting a similarity between the distant, classical bursts and the nearby
ones would have to be reassessed.

We thank the anonymous referee for his/her very useful comments.
We thank Davide Lazzati for his comments.
DG acknowledges the NSF grant AST-0307502 for financial support.

\end{document}